\documentstyle[amsmath,amssymb,12pt]{article}

\newcommand\ad{$AdS_2$}
\newcommand\adst{$AdS_2{\times}S^2$}
\newcommand\ha{\frac{1}{2}}
\newcommand\half{\textstyle{\frac{1}{2}}}

\newcommand\Psib{{\overline{\Psi}}}

\newcommand\comment[1]{{\em[#1]}}
\newcommand\skipthis[1]{{}}

\begin{document}

\rightline{HUTP-99/A031}
\rightline{hep-th/9906056}
\vskip 2cm
\centerline{\LARGE \bf Supergravity spectrum on \adst}
\vskip 2cm

\renewcommand{\thefootnote}{\fnsymbol{footnote}}
\centerline{{Jeremy
Michelson${}^{1,2}$\footnote{{\tt jeremy@yukawa.harvard.edu}.
Address after Sept.\ 1:
New High Energy Theory Center;
Rutgers University;
126~Frelinghuysen Road;
Piscataway, NJ~~08854.
}
and Marcus Spradlin${}^{2}$\footnote{\tt spradlin@feynman.harvard.edu}}} 
\vskip .5cm
\centerline{${}^1$\it Department of Physics, University of California}
\centerline{\it Santa Barbara, CA 93106}
\vskip .5cm
\centerline{${}^2$\it Department of Physics, Harvard University}
\centerline{\it Cambridge, MA 02138}
\vskip 1cm

\setcounter{footnote}{0}
\renewcommand{\thefootnote}{\arabic{footnote}}
\renewcommand{\theequation}{\thesection.\arabic{equation}}
\begin{abstract}
The Kaluza-Klein spectrum of ${\mathcal{N}}
{=}2$, $D{=}4$ supergravity compactified
on \adst\ is found and 
shown to consist of two infinite towers of
$SU(1,1|2)$ representations.
In addition to `pure gauge' modes living on the boundary
of $AdS$ which are familiar from higher dimensional
cases,
in two dimensions there are modes ({\em e.g.} massive gravitons)
which
enjoy no gauge symmetry yet nevertheless have
no on-shell degrees of freedom in the bulk.  We discuss these
two-dimensional subtleties in detail.
\end{abstract}

\newpage

\section{Introduction}
\setcounter{equation}{0}

Following the conjecture of Maldacena \cite{juan,gkp,witten}
relating supergravity (or string theory) on $AdS_d$ to a
($d{-}1$)-dimensional conformal field theory
(see \cite{huge} for a review) there has been a revival of
interest in supergravity
compactifications on anti-de Sitter spaces.
The Kaluza-Klein spectrum of these compactifications should be
identical to the spectrum of chiral primary operators in the dual
conformal field theory.
Recent papers have considered supergravity compactifications
on $AdS_3{\times}S^3$ \cite{sezgin,deboer},
$AdS_3{\times}S^2$ \cite{larsen,fujii2},
and $AdS_2{\times}S^3$ \cite{fujii3}.
We complete the square by considering
${\mathcal{N}}{=}2$, $D{=}4$ supergravity on
\adst.
Interest in \adst\ stems from the fact that it is the near-horizon
geometry of the extreme Reissner-Nordstr\"{o}m black hole in four
dimensions~\cite{frag}, and because a precise elucidation of the
nature
of the dual $0+1$ dimensional conformal field theory ({\em i.e.}
quantum
mechanics) living on the boundary of \ad\ is lacking.
For relevant work in this direction, see
\cite{andy,frag,gt}.

Compactification of supergravity to two dimensions presents
difficulties not present in higher-dimensional cases.
One is that the equation of motion for the two-dimensional
graviton $h_{\mu \nu}$ cannot be diagonalized because it involves
some scalar fields which cannot be eliminated.
However, far from being a problem, this is to be expected
in two dimensions\skipthis{.
Typically, under dimensional reduction, one obtains a
dilaton.  Usually one therefore rescales the metric so as to
remain in the Einstein frame in the lower dimension.  In $d$
dimensions this is accomplished by shifting the spacetime graviton
$h_{\mu \nu}$ by $\frac{1}{d - 2} {h^m}_m$, where ${h^m}_m$ is the
trace of the metric on the internal space (this is the linearized
Weyl shift of \cite{krv}).  However in two dimensions the
Einstein-Hilbert
action is conformally invariant and no rescaling of the metric
will eliminate the dilaton factor.}%
---conformal invariance in two dimensions prevents the definition of the
Einstein frame.
This presents no obstacle
to determining the mass spectrum, since although these
scalar fields
act as sources for the two-dimensional tensor field $h_{\mu \nu}$
they do not affect its mass.

Under dimensional reduction on $AdS$ times spheres one obtains
infinite towers of Kaluza-Klein modes with on-shell degrees of
freedom in the bulk, and in addition one typically finds a small
number of
`pure gauge' modes which live only on the boundary of $AdS$.
These include the doubleton multiplet of $SU(2,2|4)$ in the case
of $AdS_5$~\cite{krv} and the singleton multiplet of $OSp(4|8)$ in the case
of $AdS_4$~\cite{duff}.  These modes arise from fields which can be
gauged
away everywhere in the bulk using residual gauge transformations
(we discuss these
in detail in section~\ref{sec:gauge}).
In compactification to two dimensions one encounters a third type of
field:
those which have no on-shell degrees of freedom in the bulk not
because of
any gauge invariance but simply because their equations of motion
impose
sufficiently many constraints to eliminate the field completely.
In two dimensions,
massive gravitons and gravitini fall into this category.
A massive graviton, for example, enjoys no gauge invariance and has
$\half (d{+}1)(d{-}2)$ degrees of freedom on-shell in $d$ dimensions.
By a careful analysis of the equations of motion for these massive
gravitons
in \ad\ we will show that they do not give rise to independent
degrees
of freedom.

In section~\ref{sec:solution} we discuss the \adst\ solution of
${\mathcal{N}}{=}2$, $D{=}4$ pure supergravity.
In section~\ref{sec:bosons} we determine the spectrum of bosonic
Kaluza-Klein
fluctuations around the
\adst\ background.  Section~\ref{sec:gauge} contains a detailed
discussion of the residual gauge transformations in \adst.
The results of section~\ref{sec:bosons}
are fairly simple but great care is
taken to address in detail the subtleties mentioned above.
Section~\ref{sec:fermions}
addresses the fermionic Kaluza-Klein fluctuations.
Finally, in section~\ref{sec:groups} we arrange the
Kaluza-Klein modes into representations of the supergroup
$SU(1,1|2)$.

\newpage

\section{The \adst\ supergravity solution}
\label{sec:solution}
\setcounter{equation}{0}

\subsection{Lagrangian}

The gravity multiplet of ${\mathcal{N}}{=}2$, $D{=}4$ supergravity contains the
spin-$2$ graviton $g_{MN}$, a complex spin-$\frac{3}{2}$ gravitino
$\Psi_M$, and a spin-$1$ graviphoton $A_M$.  Our notational
conventions are summarized in Appendix A.  The Lagrangian
is~\cite{freedas}
\begin{equation}
\label{eq:lag}
{\cal L} = \sqrt{-g} \left[ R - \frac{1}{4} F^2 - \frac{i}{2} \Psib_M
\Gamma^{MNP} \nabla_N \Psi_P - \frac{i}{4} \Psib_M (F^{MN} - i
\Gamma^5 *\!F^{MN}) \Psi_N \right] + {\cal L}_4,
\end{equation}
where ${\cal L}_4$ contains terms quartic in $\Psi_M$ which will not
be relevant for our analysis since we are only interested in
fluctuations around a classical background with $\Psi_M = 0$.%
\footnote{In particular, we do not explore the consistency of a truncation
to the massless sector of the two-dimensional theory on \ad.} \
The
action is invariant under ${\mathcal{N}}{=}2$
supersymmetry transformations, whose
action on the gravitino field is
\begin{equation}
\label{eq:gvar}
\delta_\epsilon \Psi_M = 2 \left[ \nabla_M + \frac{i}{8} F_{NP}
\Gamma^{NP} \Gamma_M + \cdots \right] \epsilon,
\end{equation}
where $\epsilon$ is a complex spinor parameter and the dots represent
terms bilinear in $\Psi_M$ which we will not need.  We will make
extensive use of the operator
\begin{equation}
D_M \equiv \nabla_M + \frac{i}{8} F_{NP} \Gamma^{NP} \Gamma_M.
\end{equation}

\subsection{Freund-Rubin ansatz}

The bosonic equations of motion arising from~(\ref{eq:lag}) in a
background with $\Psi_M = 0$ are the familiar Einstein-Maxwell
equations
\begin{subequations}
\begin{gather}
R_{MN} - \half R g_{MN} = \half \left( F_{MP} {F_N}^{P} -
{\textstyle{\frac{1}{4}}} g_{MN} F^2 \right),\\
\nabla_M F^{MN} = 0.
\end{gather}
\end{subequations}
It follows from the trace of Einstein's equation that $R = 0$.  We
take the Freund-Rubin ansatz \cite{freund}
\begin{align}
\label{eq:freru}
F_{\mu \nu} &= 0 = F_{\mu m},& F_{mn} &= 2 \epsilon_{mn},
\end{align}
Assuming a product space, the Einstein equations then give
\begin{align}
\label{eq:frerut}
R_{\mu m} &= 0, & R_{\mu \nu} &= -g_{\mu \nu}, & R_{mn} &=
g_{mn},
\end{align}
which is consistent with the ansatz, and gives the desired
\adst\ topology, with unit radii.  

\subsection{Supersymmetry}

The \adst\ solution preserves the maximal amount of supersymmetry.
The supersymmetry variation of the bosonic fields involves only
$\Psi_M$, and consequently vanish automatically in the \adst\ 
background $\Psi_M = 0$.  The supersymmetry variation
$\delta_\epsilon \Psi_M$ is given by~(\ref{eq:gvar}) and vanishes in
the \adst\ background for Killing spinors $\epsilon$ satisfying
\begin{equation}
\label{eq:ksp}
D_M \epsilon = \left[ \nabla_M - \half \gamma \Gamma_M \right]
\epsilon = 0.
\end{equation}
The Killing spinors may be written down exactly \cite{fujii,lpr},
but we will be content to note that the integrability conditions
\begin{subequations}
\begin{gather}
0 = [ D_\mu, D_\nu ] \epsilon = \left[ {\textstyle{\frac{1}{4}}}
g^{\rho \sigma} R_{\rho \sigma}
 + \half  \right]
\gamma_{\mu\nu}
\epsilon,\\
0 = [ D_m, D_n ] \epsilon = \left[ {\textstyle{\frac{1}{4}}}
g^{pq}
R_{pq}
 - \half  \right]\gamma_{mn} \epsilon
\end{gather}
\end{subequations}
are consistent with~(\ref{eq:frerut}), implying the existence of the
maximal number of solutions for $\epsilon$, namely eight (four
complex components).

\section{Bosonic Kaluza-Klein spectrum}
\label{sec:bosons}
\setcounter{equation}{0}

In this section we calculate the spectrum of bosonic fluctuations
around the $AdS_2\times S^2$ background~(\ref{eq:freru}) and~(\ref{eq:frerut}).
We write the fluctuations as
\begin{align}
\delta g_{MN} &= h_{MN}, & \delta A_M &= a_M
\end{align}
and expand the fields in spherical harmonics $Y_{(lm)}$ on
$S^2$, which satisfy
\begin{equation}
\Box_y Y_{(lm)}(y) = -l(l+1) Y_{(lm)}(y),
\end{equation}
$(\Box_y \equiv \nabla^m\nabla_m)$.
The conventionally defined $Y_{(lm)}$ also satisfy the
useful identity
\begin{equation}
\label{eq:cute}
\nabla_m \nabla_n Y_{(1m)} = -g_{mn} Y_{(1m)}.
\end{equation}
We use parentheses to distinguish the $SO(3)$ quantum numbers
$lm$ from $S^2$ indices.
The Kaluza-Klein
modes will carry $SU(2)$ representation 
labels $(lm)$ and in particular the bosonic modes will fall into
$(2l+1)$-tuplets of $SO(3)$ with charge $l$.  Since the rank of
$SO(3)$ is one, all higher tensor harmonics can be expressed in terms
of the scalars $Y_{(lm)}$.
In particular, the space of vector spherical
harmonics is spanned by\footnote{We have not
normalized these basis vectors since we only use their orthogonality.
However, the interested reader will find the
normalization, and additional details, in~\cite{mathews}.
}
\begin{align}
\label{eq:vecylm}
\nabla_m Y_{(lm)},&& \epsilon_{mn} \nabla^n Y_{(lm)}, && l &\ge 1,
\end{align}
where we remark
for later use
that for $l{=}1$,
$K^m_{(m)}{=}\epsilon^{mn} \nabla_n Y_{(1m)}$
are the
${\mathbf 3}$ of Killing vectors on $S^2$ and
$C^m_{(m)}{=}\nabla^m Y_{(1m)}$ are the ${\mathbf 3}$ of conformal Killing
vectors on $S^2$.
The proof of this assertion relies on equation~(\ref{eq:cute}).
The space of symmetric traceless tensor
spherical harmonics is spanned
by
\begin{align}
\label{eq:sttylm}
\nabla_{\{ m} \nabla_{n\}} Y_{(lm)}, && \nabla_{\{m}
\epsilon_{n\} p} \nabla^p Y_{(lm)}, && l &\ge 2.
\end{align}

Before proceeding, it is necessary to consider carefully the
process of fixing the diffeomorphism and $U(1)$ gauge invariance
of the action to eliminate unphysical gauge degrees of
freedom.
We will see that it is possible to fix the ``de Donder''
gauge
\begin{align}
\label{eq:dedonder}
\nabla^m h_{\{mn\}} =& 0, & \nabla^m h_{\mu m} &= 0
\end{align}
and the ``Lorentz'' gauge
\begin{equation}
\label{eq:lorentz}
\nabla^m a_m = 0,
\end{equation}
but we will find 
that there are residual gauge \skipthis{invariances}transformations
preserving~(\ref{eq:dedonder}) and~(\ref{eq:lorentz}), whose
\skipthis{action} behaviour
will be important below.

\subsection{Gauge transformations and gauge fixing}
\label{sec:gauge}

The transformations of $h_{MN}$ and $a_M$ under a
diffeomorphism $\xi_M(x,y)$ and gauge transformation
$\Lambda(x,y)$ are
\begin{align}
\label{eq:trans}
\delta h_{MN} &= \nabla_M \xi_N + \nabla_N \xi_M, &
\delta a_M &= \nabla_M \Lambda + \xi^N F_{NM}
+ \nabla_M(\xi^N A_N).
\end{align}

We consider first just diffeomorphisms, parametrized by $\xi_M(x,y)$.
To fix the de Donder gauge
we need $\xi_M$ to
satisfy
\begin{align}
\nabla^m h_{\{mn\}} + (\Box_y + 1) \xi_n &= 0, &
\nabla^m h_{\mu m} + \Box_y \xi_\mu + \nabla_\mu \nabla^m \xi_m &= 0.
\end{align}
Thus, we choose
\begin{align}
\label{eq:ddd}
\xi_n &= -(\Box_y + 1)^{-1} \nabla^m h_{\{mn\}}, &
\xi_\mu &=-\Box_y^{-1}(\nabla^m h_{\mu m} + \nabla_\mu \nabla^m \xi_m).
\end{align}
We remark that there is no problem inverting the required
differential
operators on $S^2$---%
the expansion of $h_{\{mn\}}$ in terms of symmetric
traceless
tensor spherical harmonics~(\ref{eq:sttylm}) begins at $l = 2$ while
the only zero modes of $\Box_y + 1$ are the $l = 1$ vector
spherical harmonics~(\ref{eq:vecylm}), and similarly
the expansion of $h_{\mu m}$ begins at $l = 1$ while the only
zero mode of $\Box_y$ is $Y_{(00)}$.

To see if additional gauge-fixing is possible, it is convenient to
expand the diffeomorphism parameter $\xi_M(x,y)$ in
spherical harmonics:
\begin{align}
\xi_\mu(x,y) &= \xi_\mu^{(lm)} (x) Y_{(lm)}(y), &
\xi_m(x,y) &= \zeta^{(lm)} (x) \nabla_m Y_{(lm)}(y) +
\xi^{(lm)}(x) \epsilon_{mn} \nabla^n Y_{(lm)}(y),
\end{align}
where on the right-hand side the repeated $(lm)$
index implies the summation
\begin{equation}
\label{eq:sumcon}
a^{(lm)} Y_{(lm)} \equiv \sum_{l=0}^\infty \sum_{m=-l}^l
a^{(lm)} Y_{(lm)}.
\end{equation}
\skipthis{\comment{Just in case the reader is unfamiliar with $SU(2)$...}}
Maintaining the gauge~(\ref{eq:dedonder})
requires 
\begin{gather}
\nabla^m (\nabla_\mu \xi_m + \nabla_m \xi_\mu) = 0 \\
\intertext{and}
\nabla^m(\nabla_m \xi_n+\nabla_n \xi_m - g_{mn} \nabla^p \xi_p) 
= (\Box_y + 1) \xi_n = 0.
\end{gather}
The latter implies that only
$\xi^{(1m)}(x)$ and $\zeta^{(1m)}(x)$ are non-zero.  Plugging this
into the former, we find that $\xi^{(00)}_\mu$ is arbitrary while
$\xi_\mu^{(1m)} = - \nabla_\mu \zeta^{(1m)}$.

Therefore, we find that the de Donder gauge~(\ref{eq:dedonder})
is maintained by residual gauge transformations of the form
\begin{subequations} \label{keepgauge}
\begin{gather}
\xi_\mu(x,y) = \frac{1}{\sqrt{4 \pi}} \xi_\mu^{(00)}(x)
- \sum_{m=-1}^1 \nabla_\mu \zeta^{(1m)}(x) Y_{(1m)}(y),\\
\xi_n(x,y) = \sum_{m=-1}^1 \left[
\zeta^{(1m)}(x) \nabla_n Y_{(1m)}(y)
+ \xi^{(1m)}(x) \epsilon_{np} \nabla^p Y_{(1m)}(y) \right].
\end{gather}
\end{subequations}
The residual gauge parameters constitute a ${\mathbf 1}$ and two
${\mathbf 3}$'s of $SO(3)$.
The ${\mathbf 1}$ corresponds to ordinary two-dimensional
diffeomorphisms
$\xi_\mu^{(00)}$
in \ad, and the two ${\mathbf 3}$'s correspond to the three
``Yang-Mills symmetries''
$\xi^{(1m)}$ and three ``conformal diffeomorphisms''
$\zeta^{(1m)}$. %
\skipthis{The two ${\mathbf 3}$'s arise respectively from the existence of
three Killing vectors $K^m_{(m)}$
and three conformal Killing vectors $C^m_{(m)}$
on $S^2$.}
We will consider the residual gauge transformations further in
section~\ref{sec:resgauge}. 

Having explored diffeomorphisms to our satisfaction, we now need to
find $\Lambda(x,y)$ so that the Lorentz gauge~(\ref{eq:lorentz})
is achieved.  This is obviously accomplished via
\begin{equation}
\label{eq:eee}
\Lambda(x,y) = -\Box_y^{-1} \nabla^m a_m.
\end{equation}
Again there is no problem inverting the Laplacian.  The residual
gauge transformations are those for which $\Box_y \Lambda = 0$,
{\em i.e.},
$\Lambda(x,y) = \Lambda(x)$.
These provide the usual two dimensional gauge invariance, as discussed
further in section~\ref{sec:resgauge}.

In this subsection we verified that it is possible to fix
the de Donder~(\ref{eq:dedonder}) and Lorentz~(\ref{eq:lorentz})
gauges, and we have found
that there are a finite set of residual gauge transformations
preserving these gauge choices.
These residual gauge transformations differ from
the gauge transformations~(\ref{eq:ddd}) and~(\ref{eq:eee}) used
to fix de Donder and Lorentz gauges in an essential
way.
Finding the latter gauge transformations
involved
inverting differential operators on $S^2$, which we showed was always
possible to do, whereas fixing the residual gauge degrees of freedom
involves
inverting differential operators on
\ad, a process which is not completely well-defined because
of ambiguity in the choice of $\Box_x^{-1}$ related to a choice
of boundary conditions at the edge of \ad.
The residual gauge transformations may be used to eliminate the bulk
degrees of freedom associated with a given field, but they do not eliminate
boundary degrees of freedom associated with that mode.
That is, the fields can be gauged away everywhere except on the
boundary
of \ad.  The modes killed by imposing the de Donder and Lorentz
gauges, on the other hand,
have no residual degrees of freedom on the
boundary and can be ignored completely.

\subsection{Spherical harmonic expansion}

We expand the bosonic fields
as
\begin{subequations}
\label{eq:fluct}
\begin{align}
h_{\mu \nu}(x,y) &= H^{(lm)}_{\mu \nu}(x) Y_{(lm)}(y),  \\
h_{\mu m}(x,y) &= B^{(lm)}_\mu(x)
\epsilon_{m n} \nabla^n Y_{(lm)}(y) +
\tilde{B}^{(lm)}_\mu(x) \nabla_m Y_{(lm)}(y), \\
h_{\{mn\}}(x,y) &= \phi^{(lm)}(x) \nabla_{\{m}\nabla_{n\}}
Y_{(lm)}(y)
+ \tilde{\phi}^{(lm)}(x)
\nabla_{\{m} \epsilon_{n\} p}\nabla^p Y_{(lm)}, \\
{h^{m}}_m(x,y) &= \pi^{(lm)}(x) Y_{(lm)}(y), \displaybreak[0] \\
a_\mu(x,y) &= b^{(lm)}_\mu(x) Y_{(lm)}(y), \\
a_m(x,y)
&= b^{(lm)}(x) \epsilon_{mn} \nabla^n Y_{(lm)}(y) +
\tilde{b}^{(lm)}(x)
\nabla_m
Y_{(lm)}(y)\skipthis{,}%
.
\end{align}
\end{subequations}
\skipthis{where we continue to use the summation convention~(\ref{eq:sumcon}).}
Imposing the de Donder~\eqref{eq:dedonder} and Lorentz~\eqref{eq:lorentz}
gauges
to fix local diffeomorphism and $U(1)$ gauge invariance eliminates
$\tilde{B}_\mu$, $\phi$, $\tilde{\phi}$, and $\tilde{b}$,
leaving
\begin{subequations}
\label{eq:boson}
\begin{align}
h_{\mu \nu}(x,y) &= H^{(lm)}_{\mu \nu}(x) Y_{(lm)}(y),  \\
h_{\mu m}(x,y) &= B^{(lm)}_\mu(x) \epsilon_{m n} \nabla^n
Y_{(lm)}(y), \\
h_{\{mn\}}(x,y) &= 0, \\
{h^m}_m(x,y) &= \pi^{(lm)}(x) Y_{(lm)}(y), \displaybreak[0] \\
a_\mu(x,y) &= b^{(lm)}_\mu(x) Y_{(lm)}(y), \\
a_m(x,y) &= b^{(lm)}(x) \epsilon_{mn} \nabla^n Y_{(lm)}(y).
\end{align}
\end{subequations}
We note for later use that
we can set
\begin{align}
\label{eq:note}
b^{(00)} &= 0, & B_\mu^{(00)} &= 0
\end{align}
since the corresponding vector spherical harmonic vanishes
at $l{=}0$.

\subsection{Residual gauge transformations} \label{sec:resgauge}

Recalling equation~\eqref{keepgauge}, we now consider separately the
action of 
each of these residual gauge transformations.

Under an ordinary diffeomorphism $\xi_\mu^{(00)}$ we have
\begin{equation}
\delta h_{\mu\nu}
=
\frac{1}{\sqrt{4 \pi}}
(\nabla_\mu \xi_\nu^{(00)} +
\nabla_\nu \xi_\mu^{(00)}),
\end{equation}
which is indeed the action of a usual two-dimensional diffeomorphism.
In particular, upon comparison with the expansion of $h_{\mu\nu}$
in~(\ref{eq:fluct}), we find
\begin{equation}
\label{eq:diffeo}
\delta H_{\mu\nu}^{(00)} = \nabla_\mu \xi_\nu^{(00)} +
\nabla_\nu \xi_\mu^{(00)}.
\end{equation}
Thus $\xi_\mu^{(00)}$ provides gauge invariance for
$H_{\mu\nu}^{(00)}$, which we will find to be the massless
\ad\ graviton.

The action of
$\xi^{(1m)}$
is
\begin{equation}
\delta h_{\mu n}
=\sum_{m=-1}^1 \nabla_\mu \xi^{(1m)} \epsilon_{np} \nabla^p Y_{(1m)},
\end{equation}
which upon comparison with the expansion of $h_{\mu n}$
in~(\ref{eq:fluct})
gives
\begin{equation}
\label{eq:Uone}
\delta B_\mu^{(1m)} = \nabla_\mu \xi^{(1m)}.
\end{equation}
Thus $\xi^{(1m)}$
provides gauge invariance for the $SO(3)$
triplet of vectors $B_\mu^{(1m)}$, which we will find to be massless.

The action of
$\zeta^{(1m)}$
on $h_{MN}$ takes the form
\begin{subequations}
\begin{gather}
\delta {h^n}_n = -4 \sum_{m=-1}^1 \zeta^{(1m)} Y_{(1m)},\\
\delta h_{\mu \nu} = -2 \sum_{m=-1}^1
\nabla_\mu \nabla_\nu \zeta^{(1m)} Y_{(1m)},
\end{gather}
\end{subequations}
which amounts to
\begin{align}
\label{eq:conformal}
\delta \pi^{(1m)} &= -4 \zeta^{(1m)}, &
\delta H_{\mu\nu}^{(1m)} &= -2 \nabla_\mu \nabla_\nu \zeta^{(1m)}.
\end{align}
We can use these conformal diffeomorphisms to set some
linear combination of ${H^{(1m) \mu}}_\mu$ and $\pi^{(1m)}$ to zero.

Also, the residual gauge transformation $\Lambda(x)$, left undefined by
equation~\eqref{eq:ddd}, gives
\begin{equation}
\label{eq:uone}
\delta b_\mu^{(00)} = \nabla_\mu \Lambda(x)
\end{equation}
for the vector field $b_\mu^{(00)}$, which we will also find
to be massless.

\subsection{Equations of motion}

The linearized Einstein equations are
\begin{subequations}
\begin{gather}
-\ha \nabla_\mu \nabla_\nu({h^\rho}_\rho + {h^m}_m) - \ha (\Box_x
+ \Box_y + 2) h_{\mu \nu} +
\nabla_{(\mu} \nabla^\rho h_{\nu)\rho}
+ g_{\mu\nu} {h^\rho}_\rho
= g_{\mu\nu} ({h^m}_m-\epsilon^{mn} \nabla_m a_n),
\\
-\ha \nabla_\mu \nabla_m ({h^\nu}_\nu + \ha {h^n}_n) - \ha
(\Box_x + \Box_y - 2) h_{\mu m} + 
\nabla_{(\mu} \nabla^\nu h_{m)\nu}
=
{\epsilon_m}^n (\nabla_\mu a_n - \nabla_n a_\mu),
\\
- \ha \nabla_m \nabla_n {h^\rho}_\rho - \ha (\Box_x + \Box_y - 2)
h_{mn} + \nabla_{(m} \nabla^\mu h_{n)\mu}
= g_{mn} \epsilon^{pq} \nabla_p
a_q,
\end{gather}
and the linearized Maxwell equations, written in components, are
\begin{gather}
0 =
-2 \epsilon^{mn} \nabla_m h_{n \mu} + (\Box_x + \Box_y + 1) a_\mu
- \nabla_\mu \nabla^\nu a_\nu,\\
0 =
2 {\epsilon_m}^n \nabla^\mu h_{\mu n} - \epsilon_{mn}
\nabla^n ({h^\mu}_\mu - {h^p}_p) + (\Box_x + \Box_y - 1) a_m
- \nabla_m \nabla^\mu a_\mu.
\end{gather}
\end{subequations}
Inserting~(\ref{eq:boson}) into these equations of motion gives
\begin{subequations}
\begin{multline}
\label{eq:one}
\Big[-
 \half \nabla_\mu \nabla_\nu({H^{(lm) \rho}}_\rho +
\pi^{(lm)}) - \half (\Box
- (l+2)(l-1)) H^{(lm)}_{\mu \nu}
+ \nabla_{(\mu} \nabla^\rho H^{(lm)}_{\nu)\rho}
\\
+ g_{\mu\nu} ({H^{(lm) \rho}}_\rho - \pi^{(lm)}
+ l(l+1) b^{(lm)})\Big] Y_{(lm)} = 0,
\end{multline}
\begin{multline}
\label{eq:two}
\Big[ -\half \nabla_\mu ({H^{(lm) \rho}}_\rho + \half \pi^{(lm)})
+ \half \nabla^\rho H^{(lm)}_{\rho \mu}
+ \nabla_\mu b^{(lm)}\Big]\nabla_m Y_{(lm)},
\\
=
\Big[ 
\half (\Box - l(l+1)-1) B^{(lm)}_\mu
-\half \nabla_\mu \nabla^\nu B^{(lm)}_\nu
- b^{(lm)}_\mu\Big] \epsilon_{mn} \nabla^n Y_{(lm)},
\end{multline}
\begin{multline}
\label{eq:three}
\Big[ - \half {H^{(lm) \rho}}_\rho \Big] \nabla_{\{m} \nabla_{n\}}
Y_{(lm)}
+  \Big[
\nabla^\mu B^{(lm)}_\mu \Big] \nabla_{\{m} \epsilon_{n\} p} \nabla^p
Y_{(lm)}\\
+ g_{mn} \Big[ {\textstyle{\frac{1}{4}}}l(l+1) {H^{(lm) \rho}}_\rho
- {\textstyle{\frac{1}{4}}}(\Box - l(l+1)-2)\pi^{(lm)} -
l(l+1)b^{(lm)}
\Big]
Y_{(lm)}
=0,
\end{multline}
\begin{gather}
\label{eq:four}
\Big[ -2 l(l+1) B^{(lm)}_\mu + (\Box - l(l+1)+1) b^{(lm)}_\mu 
- \nabla_\mu \nabla^\nu b^{(lm)}_\nu \Big] Y_{(lm)} = 0,
\\ \intertext{and}
\label{eq:five}
\Big[ -2 \nabla^\mu B^{(lm)}_\mu - \nabla^\mu b^{(lm)}
_\mu
\Big] \nabla_m Y_{(lm)} -
 \Big[ {H^{(lm) \mu}}_\mu - \pi^{(lm)} - 
(\Box - l(l+1)) b^{(lm)} \Big] \epsilon_{mn} \nabla^n Y_{(lm)} = 0.
\end{gather}
\end{subequations}
Each term in brackets must vanish separately (for each
$(lm)$) due to the orthogonality
of the various spherical harmonics.
However, not all of the equations are in effect for $l{=}0$ and
$l{=}1$.  For example, in~(\ref{eq:three}), $\nabla_{\{m}
\nabla_{n\}} Y_{(lm)}$ vanishes for $l{<}2$ so we only get the
condition ${H^{(lm)\rho}}_\rho{=}0$ for $l{\ge}2$.
For
this reason it is worthwhile to consider the $l{=}0$ and $l{=}1$
levels separately from $l{\ge}2$.
First, however, we present a brief but useful diversion
regarding massive gravitons.

\subsection{A note on massive gravitons in \ad}
\label{sec:note}

Consider a symmetric traceless tensor $k_{\mu \nu}$ in two
dimensions which is also transverse, {\em i.e.}
\begin{equation}
\label{eq:transverse}
\nabla^\mu k_{\mu\nu} = 0.
\end{equation}
In two dimensions, these assumptions are sufficient to prove
that
\begin{equation}
\label{eq:boxr}
(\Box - R) k_{\mu \nu} = 0,
\end{equation}
where $R$ is the
two-dimensional
Ricci scalar, with $R = -2$ for \ad.
Now suppose that $k_{\mu \nu}$ is to satisfy the equation of
motion for a graviton of mass $m^2$
in $AdS$ \cite{polishchuk}
\begin{equation}
\label{eq:geom}
(\Box + 2 - m^2) k_{\mu\nu} - 2 \nabla_{(\mu } \nabla^\rho k_{\nu)
\rho} = 0.
\end{equation}
The second term vanishes by virtue of~(\ref{eq:transverse}), and
the remaining equation is consistent with~(\ref{eq:boxr}) only
if $m^2 = 0$.
We conclude that a symmetric traceless tensor field $k_{\mu \nu}$
in \ad\ which satisfies~(\ref{eq:transverse}) can only
satisfy~(\ref{eq:geom})
if $m^2 = 0$.  If $m^2 \ne 0$, the only solution
is $k_{\mu\nu} = 0$.
We will use this analysis in the following sections to eliminate
the massive graviton from the Kaluza-Klein spectrum.

\subsection{The $l = 0$ level}

Recalling~(\ref{eq:note}), we only have to
worry about the fields $H_{\mu\nu}^{(00)}$, $\pi^{(00)}$, and
$b_\mu^{(00)}$.
There are only three equations, coming from the brackets
multiplying $Y_{(00)}$ in~(\ref{eq:one}),~(\ref{eq:three}),
and~(\ref{eq:four}):
\begin{subequations}
\label{eq:aaa}
\begin{gather}
0 =
- \ha \nabla_\mu\nabla_\nu ({H^{(00) \rho}}_\rho
+ \pi^{(00)}) - \ha (\Box+2) H^{(00)}_{\mu\nu}
+ \nabla_{(\mu} \nabla^\rho H^{(00)}_{\nu)\rho} +
g_{\mu \nu} ({H^{(00) \rho}}_\rho - \pi^{(00)}),\\
0=
-\frac{1}{4}(\Box - 2) \pi^{(00)},\\
0 =
(\Box - 1) b_\mu^{(00)}  - \nabla_\mu \nabla^\nu b_\nu^{(00)}.
\end{gather}
\end{subequations}
The second equation gives a scalar with mass $m^2 = 2$, while the
third equation may be written as
\begin{equation}
\nabla^\nu f_{\mu \nu} = 0, ~~~~~ f_{\mu \nu}
= \nabla_\mu b_\nu^{(00)} - \nabla_\nu b_\mu^{(00)},
\end{equation}
(the shift of the Laplacian by $-1$ comes from the curvature of
$S^2$) so $b_\mu^{(00)}$ is a massless vector, which has the
appropriate gauge invariance as a result of~(\ref{eq:uone}).
In two dimensions
a massless vector has no propagating degrees of freedom on shell,
but there are boundary degrees of freedom which we will tabulate
below.

It remains to analyze the third equation in~(\ref{eq:aaa}).
There is no constraint on the trace of $H_{\mu\nu}^{(00)}$ so
we will consider the trace part and the traceless part separately.
First consider turning on a fluctuation with $\pi^{(00)} = 0$
and $H^{(00)}_{\mu \nu} = H g_{\mu \nu}$.
The equation of motion for such a mode is
\begin{equation}
(\Box - 2)H = 0,
\end{equation}
which seems to give a scalar field $H$ with mass $m^2 = 2$.
However, one can use the residual diffeomorphism
invariance~(\ref{eq:diffeo})
to gauge $H$ to zero everywhere except on the boundary.
Therefore $H$ is
not a propagating degree of freedom in the bulk but is
present on the boundary.

Now consider the traceless part of $H^{(00)}_{\mu\nu}$.
The equation of motion for a fluctuation $H^{(00)}_{\mu\nu}$
with ${H^{(00) \rho}}_\rho = 0$ is then
\begin{equation}
\label{eq:hhh}
(\Box + 2) H_{\mu \nu}^{(00)} -
2 \nabla_{(\mu} \nabla^\rho H^{(00)}_{\nu) \rho}
= -\nabla_\mu\nabla_\nu
 \pi^{(00)}
- 2 g_{\mu\nu}\pi^{(00)}.
\end{equation}
As discussed in the introduction, the scalar field $\pi^{(00)}$
cannot be eliminated from this equation.
Nevertheless, the source terms on the right-hand
side of~(\ref{eq:hhh}) do not affect the mass of $H_{\mu\nu}^{(00)}$.
For $\pi^{(00)} = 0$,~(\ref{eq:hhh}) is precisely
the equation of motion for a massless graviton~(\ref{eq:geom}),
and indeed $H_{\mu\nu}^{(00)}$ has
the correct \ad\ diffeomorphism invariance to be the massless
graviton as a
consequence of~(\ref{eq:diffeo}).
Because of this gauge invariance, $H_{\mu\nu}^{(00)}$ can be
eliminated in the bulk, and the massless graviton has only
a boundary degree of freedom.

To summarize: at the $l = 0$ level we find one bulk degree
of freedom, a scalar with $m^2 = 2$, and three pure gauge modes:
a scalar with $m^2 = 2$, a massless vector, and a massless graviton.

\subsection{The $l = 1$ level}

At the $l = 1$ level there are seven equations, coming from
all of the brackets in~(\ref{eq:one})--(\ref{eq:five}) except
for the first two in~(\ref{eq:three}).
From~(\ref{eq:three}) and~(\ref{eq:five}) we find
\begin{subequations}
\label{eq:further}
\begin{gather}
0 = \half {H^{(1m) \rho}}_\rho - {\textstyle{\frac{1}{4}}}(\Box
-4)\pi^{(1m)} - 2 b^{(1m)},\\
0 = {H^{(1m) \rho}}_\rho - \pi^{(1m)} - (\Box - 2)b^{(1m)},
\end{gather}
\end{subequations}
which combine to yield
\begin{equation}
(\Box - 6)(\pi^{(1m)} - 2 b^{(1m)}) = 0,
\end{equation}
so that $\pi^{(1m)} - 2 b^{(1m)}$ is a scalar with $m^2 = 6$.
Since we can use the residual conformal
diffeomorphisms~(\ref{eq:conformal})
to gauge-fix $\pi^{(1m)}$
to zero, there is no other scalar at this level.
Since the conformal diffeomorphisms eliminate only
bulk degrees of freedom, there is an extra boundary
degree of freedom associated to an $l = 1$ ``scalar.''
However, because the field equation for
this excitation is missing, we cannot determine its mass (although in
section~\ref{sec:groups} we will use group theory to argue that it
must have $m^2 = 0$).
The missing equation of motion and the use of the residual
conformal diffeomorphisms is discussed by \cite{krv} in the context
of $AdS_5 \times S^5$, where the boundary degrees of freedom in
that case turn out to be the six scalars of the $SU(2,2|4)$
doubleton.

The equations~(\ref{eq:further}) show that ${H^{(1m) \rho}}_\rho$
is not an indepedent degree of freedom
since it can be algebraically eliminated.
Furthermore, one also finds from~(\ref{eq:one})
and~(\ref{eq:two})
\begin{subequations}
\begin{gather}
\label{eq:mmm}
\Box H_{\mu\nu}^{(1m)} - 2
\nabla_{(\mu} \nabla^\rho H^{(1m)}_{\nu)\rho}
= -\nabla_\mu \nabla_\nu
({H^{\rho (1m)}}_\rho+\pi^{(1m)}) + 2 g_{\mu\nu}({H^{\rho (1m)}}_\rho
-\pi^{(1m)}
+ 2 b^{(1m)}), \\
\label{eq:ggg}
0 = -\half \nabla_\mu ( {H^{(1m) \rho}}_\rho
+ \half \pi^{(1m)}) + \half \nabla^\rho H^{(1m)}_{\rho \mu}
+ \nabla_\mu b^{(1m)}.
\end{gather}
\end{subequations}
The second of these equations
gives two first order constraints on $H^{(1m)}_{\mu\nu}$,
enabling one to solve for the remaining two components.
Thus $H_{\mu\nu}^{(1m)}$ has no on-shell degrees of freedom in the
bulk,
as we expect for gravitons in two dimensions.
To be more precise, consider a traceless fluctuation of
$H_{\mu\nu}^{(1m)}$.  The linear differential
equations~(\ref{eq:ggg})
will have a specific solution involving $b^{(1m)}$ and $\pi^{(1m)}$,
as well as a homogeneous solution $\hat{H}_{\{\mu\nu\}}^{(1m)}$
satisfying
\begin{equation}
\label{eq:homogen}
\nabla^\rho \hat{H}_{\{\rho \mu\}}^{(1m)} = 0.
\end{equation}
But the analysis of section~\ref{sec:note} (with $m^2{=}l(l{+}1){=}2$) shows that
$\hat{H}_{\{\rho \mu\}}^{(1m)}$ must vanish,
implying that $H_{\{\mu \nu\}}^{(1m)}$ has no independent degrees
of freedom but is completely determined in terms of
$b^{(1m)}$ and $\pi^{(1m)}$ through the specific solution
to~(\ref{eq:ggg}).

Finally, from~(\ref{eq:two}),~(\ref{eq:four}), and~(\ref{eq:five})
we have the three equations
\begin{subequations}
\begin{gather}
0 = \half (\Box-3) B_\mu^{(1m)} - \half \nabla_\mu \nabla^\nu
B^{(1m)}_\nu-b^{(1m)}_\mu,\\
0 = -4 B^{(1m)}_\mu + (\Box-1)b^{(1m)}_\mu - \nabla_\mu
\nabla^\nu b^{(1m)}_\nu,\\
0 = -2 \nabla^\mu B^{(1m)}_\mu - \nabla^\mu b^{(1m)}_\mu.
\end{gather}
\end{subequations}
We can use the gauge invariance~(\ref{eq:Uone}) of $B^{(1m)}_\mu$
to
set $\nabla^\mu B^{(1m)}_\mu = 0$, which we see
then sets $\nabla^\mu b^{(1m)}_\mu
= 0$ as well.  The remaining two equations decouple to give
\begin{equation}
(\Box-5)(b_\mu^{(1m)} + B_\mu^{(1m)}) = 0, ~~~~~
(\Box + 1)(2 b_\mu^{(1m)} -B_\mu^{(1m)}) = 0,
\end{equation}
which are the equations of motion for vectors of mass
$m^2 = 6$ and $m^2 = 0$, respectively.
The massless vector has the proper gauge invariance as
a result of~(\ref{eq:Uone}).

We have exhausted the content of all seven equations.
To summarize, at the $l = 1$ level we find two bulk
degrees of freedom, one scalar and one vector each with
$m^2 = 6$, and boundary degrees of freedom corresponding
to a scalar whose mass we could not determine and one massless
vector.  Each of these modes is a ${\mathbf 3}$ of
$SO(3)$.

\subsection{$l \ge 2$}

For $l \ge 2$
we immediately observe from~(\ref{eq:three}) that
${H^{(lm) \rho}}_\rho = 0 = \nabla^\mu B^{(lm)}_\mu$, and hence,
from~(\ref{eq:five}), that $\nabla^\mu
b^{(lm)}_\mu = 0$.
The two equations involving only the vectors $B^{(lm)}_\mu$ and
$b^{(lm)}_\mu$
come from~(\ref{eq:two}) and~(\ref{eq:four}) and
now read
\begin{equation}
(\Box+1)
\begin{pmatrix}
B_\mu^{(lm)} \\
b_\mu^{(lm)}
\end{pmatrix}
= 
\begin{pmatrix}
l(l+1)+2 & 2\\
2l(l+1) & l(l+1)
\end{pmatrix}
\begin{pmatrix}
B_\mu^{(lm)} \\
b_\mu^{(lm)}
\end{pmatrix}
\end{equation}
Diagonalizing this mass matrix leads to two infinite towers of
massive vector fields with on-shell masses $m^2 = \Box + 1$ (where
the shift comes from a curvature term in the Maxwell equations)
given by
\begin{equation}
m^2 = \left\{ \begin{matrix}
l(l-1)\\
(l+1)(l+2)
\end{matrix}
\right., \qquad l \ge 2.
\end{equation}
The two equations involving only the scalars $\pi^{(lm)}$
and $b^{(lm)}$
coming from~(\ref{eq:three}) and~(\ref{eq:five})
are
\begin{equation}
\Box \begin{pmatrix} \pi^{(lm)} \\
b^{(lm)}
\end{pmatrix}
 = 
\begin{pmatrix}
l(l+1) +2 & -4l(l+1)\\
-1 & l(l+1)
\end{pmatrix}
\begin{pmatrix} \pi^{(lm)} \\
b^{(lm)}
\end{pmatrix}
,
\end{equation}
and hence we have two towers of massive scalars with masses
($m^2 = \Box$)
\begin{equation}
m^2 = \left\{
\begin{matrix}
l(l-1)\\
(l+1)(l+2)
\end{matrix}\right. , \qquad l \ge 2.
\end{equation}
We have exhausted the content of all but two equations.
From~(\ref{eq:one}) and~(\ref{eq:two}) we have
\begin{subequations}
\label{eq:bbb}
\begin{gather}
(\Box - (l+2)(l-1)) H_{\mu \nu}^{(lm)}
- 2 \nabla_{(\mu} \nabla^\rho
H_{\nu)\rho}^{(lm)} = - \nabla_\mu \nabla_\nu \pi^{(lm)}
+ 2 g_{\mu \nu}(-\pi^{(lm)} + l(l+1) b^{(lm)}),
\\
-\frac{1}{4} \nabla_\mu \pi^{(lm)} + \ha \nabla^\rho
{H^{(lm)}_{\rho \mu}} + \nabla_\mu b^{(lm)} = 0.
\end{gather}
\end{subequations}
We now mimick the analysis following~(\ref{eq:mmm})
and~(\ref{eq:ggg})
to conclude that $H_{\mu\nu}^{(lm)}$ has no on-shell
degrees of freedom.

\subsection{Summary}

We separate the bosonic spectrum into the physical (propagating)
degrees
of freedom in the bulk and those which live only on
the boundary.  The bulk modes are
\begin{equation}
\label{eq:bosmas}
\begin{tabular}{|c|c|c|}\hline
scalar&$m^2 = l(l-1)$&$l \ge 2$\\ \hline
scalar&$m^2 = (l+1)(l+2)$&$l \ge 0$\\ \hline
vector&$m^2 = l(l-1)$&$l \ge 2$\\ \hline
vector&$m^2 = (l+1)(l+2)$&$l\ge 1$\\ \hline
\end{tabular}
\end{equation}
where each field represents one on-shell bosonic degree of freedom.
The modes which live only on the boundary are
\begin{equation}
\label{eq:bosmas2}
\begin{tabular}{|c|c|c|}\hline
scalar & $m^2 = \ ?$ & $l = 1$\\ \hline
scalar & $m^2 = 2$ & $l = 0$\\ \hline
vector & $m^2 = l(l-1) = 0$ & $l = 0,1$\\ \hline
graviton & $m^2 = 0$ & $l = 0$\\ \hline
\end{tabular}
\end{equation}
The first scalar has no associated field equation so it is impossible
to determine its mass.
(In section~\ref{sec:groups} we will use group theory
to argue that the mass of this scalar is $m^2 = 0$.)
All of the entries in this table represent
pure gauge modes, {\em i.e.} the corresponding field can be
gauged away everywhere except
on the boundary of \ad\ using the residual gauge invariances
disussed in section~\ref{sec:gauge}.

The tables~(\ref{eq:bosmas}) and~(\ref{eq:bosmas2}) are not very
useful.
For one thing, the distinction between scalars and vectors makes
no sense in two dimensions, where each one has one on-shell degree
or freedom.   Secondly,
the $m^2$ column doesn't really make sense in
{\ad}--- on-shell, $m^2$ is the quadratic Casimir of the Poincar\'e
group, which is the isometry group of flat space but not of \ad.
Instead of $m^2$, we should label our bosonic modes by their
quantum numbers $(h,q)$ under the $SL(2,{\Bbb R}) \times SU(2)$
isometry group of \adst.

For bosonic fields in \ad, the conformal weight $h$ is related to
the ``mass'' (as conventially defined) by the well-known
formula \cite{huge}
\begin{equation}
h_\pm = \half (1 \pm \sqrt{1 + 4 m^2}),
\end{equation}
where one takes the larger root.  The $SU(2)$ charge $q$ is
simply the $l$ of the associated spherical harmonic $Y_{(lm)}$.
Using these
rules we find that the $SL(2,{\Bbb R}) \times SU(2)$
quantum numbers of the bulk modes are
\begin{equation}
\label{eq:groupbosons}
(2,0), ~~~ 2(k,k)_{k \ge 2}, ~~~ 2(k+1,k-1)_{k \ge 2},
\end{equation}
and those of the boundary modes are
\begin{equation}
\label{eq:groupgauge}
(?,1), ~~~ (2,0), ~~~ (1,1), ~~~ 2 (1,0).
\end{equation}
In section~\ref{sec:groups} we argue that the first mode must
be $(1,1)$.

\section{Fermionic Kaluza-Klein spectrum}
\label{sec:fermions}
\setcounter{equation}{0}

In this section we calculate the spectrum of fermionic fluctuations
around the \adst\ background $\Psi_M{=}0$.
The Lagrangian~(\ref{eq:lag})
gives the following equation of motion for
$\Psi_M$ in a background with $\Psi_M{=}0$,
\begin{equation}
\Gamma^{MNP} \nabla_N \Psi_P + {\textstyle{\frac{i}{2}}}
(F^{MN} - i \Gamma^5 *\!F^{MN})
\Psi_N = 0.
\end{equation}
In the \adst\ background given
by~(\ref{eq:freru}) and~(\ref{eq:frerut}) this equation
takes the simple form
\begin{equation}
\label{eq:feomt}
\Gamma^{MNP} D_N \Psi_P = 0,
\end{equation}
with $D_N$ given by~(\ref{eq:ksp}).
We decompose the four-dimensional gravitino $\Psi_M(x,y)$ into
$\Psi_\mu(x,y)
= \psi_\mu(x) \otimes \chi(y)$ and $\Psi_m(x,y) = \psi(x)
\otimes
\chi_m(y)$, where
$\psi_\mu(x)$ and $\psi(x)$ are spacetime gravitini and
spinors and $\chi(y)$ and $\chi_m(y)$ are internal ($S^2$)
spinors and vector-spinors.
The equation of motion~(\ref{eq:feomt})
couples the spacetime gravitini and
spinor degrees of freedom, but they can be decoupled by imposing
the condition
\begin{equation}
\label{eq:fgauge}
\Gamma^M \Psi_M = 0
\end{equation}
to gauge-fix
local supersymmetry.  Using the equation of motion~(\ref{eq:feomt}),
one finds that the condition~(\ref{eq:fgauge}) also implies that
\begin{equation}
\label{eq:gauget}
D_M \Psi^M = \nabla_M \Psi^M = 0.
\end{equation}
Using the conditions~(\ref{eq:fgauge})
and~(\ref{eq:gauget}), the decoupled equations
of motion for $\psi_\mu$ and $\psi$ read
\begin{subequations}
\label{eq:feom}
\begin{gather} \label{eq:feomgrav}
(
\gamma^\nu
\nabla_\nu \psi^\mu- \gamma^{\mu\nu}\psi_\nu )\otimes \chi =
\psi^\mu \otimes (- \gamma
\gamma^m \nabla_m \chi),\\
\label{eq:feomspin}
( \gamma^\mu \nabla_\mu \psi) \otimes \chi^m = \psi \otimes
(- \gamma \gamma^n \nabla_n \chi^m -  \gamma^{mn} \chi_n).
\end{gather}
\end{subequations}

\subsection{Spinor harmonics on $S^2$}

It will be convenient to expand the fermionic fields in terms of
eigenspinors of the `Dirac' operator $-\gamma \gamma^m \nabla_m$
on $S^2$.  We denote the eigenspinors of this operator
by $\eta_{(lm)}(y)$, where $l$ is now a half-integer, $l = \half,
{\textstyle{\frac{3}{2}}},\ldots$, and $m$ ranges as
as usual from $-l$ to $+l$ in unit steps.  These spinors
satisfy
\begin{equation}
-\gamma \gamma^m \nabla_m \eta_{(lm)}(y) = (l + \half) \eta_{(lm)}(y),
\end{equation}
and hence
\begin{equation}
\Box_y \eta_{(lm)}(y) = \left[-l(l+1) + {\textstyle{1 \over 4}}\right]
\eta_{(lm)}(y).
\end{equation}
Note that $\eta_{(\ha,m)}$ are the ${\bf 2}$ of Killing spinors on
$S^2$ satisfying
\begin{equation}
D_m \eta_{(\ha,m)} = \left[\nabla_m - \half \gamma \gamma_m \right]
\eta_{(\ha,m)} = 0,
\end{equation}
and indeed the $\eta_{(lm)}$'s may be constructed from 
$Y_{(lm)}$ and $\eta_{(\ha,m)}$ \cite{krv}.

\subsection{Spinors}

The spectrum of spacetime spinors $\psi(x)$ is found by diagonalizing the
mass
operator
\begin{equation}
M^{(1/2)}_{mn} \equiv -g_{mn} \gamma
\gamma^p \nabla_p - \gamma_{mn}
\end{equation}
on $S^2$.
A complete basis of vector-spinors $\chi_m$ on $S^2$ is provided by
\begin{gather}
\label{eq:basvs}
\begin{align} 
\gamma_m \eta_{(lm)}, &&
\gamma_m \gamma \eta_{(lm)}, &&
\nabla_m \eta_{(lm)}, &&
\nabla_m \gamma \eta_{(lm)}.
\end{align}
\end{gather}
However, this basis includes modes of the form $\chi_m
= D_m \chi$ (for some $\chi$),
which are spurious since they can be gauged away
using the gauge symmetry~(\ref{eq:gvar}).
Using the definitions of $D_m$ and $M_{mn}^{(1/2)}$, one can show
that
\begin{equation}
\label{eq:showt}
D^m M_{mn}^{(1/2)} \chi^n = -\gamma \gamma^n \nabla_n D_m \chi^m.
\end{equation}
If one now decomposes $\chi_m$ into \cite{cenr}
\begin{align}
\chi_m &= \varphi_m + D_m \varphi, & D^m \varphi_m &= 0,
\end{align}
then~(\ref{eq:showt}) imples that this decomposition is maintained by
the spin-$\ha$ mass operator.  Therefore, the eigenspace of this
operator
naturally decomposes into the space of spurious states with
$D_m \chi^m
\ne 0$ and the space of physical states which obey
\begin{equation}
D^m \chi_m = 0.
\end{equation}
Viewing $D^m$ as an operator from the four-dimensional space of
vector-spinors spanned
by~(\ref{eq:basvs}) (at fixed $l$ and $m$)
to the two-dimensional space of spinors spanned
by $\eta_{(lm)}$ and $\gamma \eta_{(lm)}$,
it is clear that
$D^m$ has (at least) a two-dimensional kernel.
One can check that the kernel is exactly two-dimensional and that a
basis
of the physical states satisfying $D^m \chi_m = 0$ is provided by
the two vector-spinors
\begin{subequations}
\begin{align}
A^m_{(lm)} &= 
\gamma \left[ \nabla^m - l \gamma \gamma^m \right] \eta_{(lm)}, &
l \ge {\textstyle{3 \over 2}},\\
B^m_{(lm)} &=
\left[ \nabla^m + (l+1) \gamma^m \gamma \right] \eta_{(lm)},
&  l \ge \half,
\end{align}
\end{subequations}
where the restriction on $l$ for $A^m$ arises from noting that
$A^m_{(\ha,m)} = \gamma D^m \eta_{(\ha,m)} = 0$.
In this basis $M_{mn}^{(1/2)}$ acts diagonally,
\begin{equation}
\label{eq:smatrix}
M_{mn}^{(1/2)} \begin{pmatrix}
A^n_{(lm)}\\
B^n_{(lm)}
\end{pmatrix}
= \begin{pmatrix}
(l + \half) A_{m (lm)}\\
-(l + \half) B_{m (lm)}
\end{pmatrix}.
\end{equation}

Let us summarize
$SL(2,{\Bbb R})
\times SU(2)$ content of these spinors.
For each eigenvector-spinor $\chi_m$ of $M^{(1/2)}_{mn}$
with eigenvalue $\lambda$
there will be an \ad\ spinor $\psi$  satisfying~(\ref{eq:feomspin})
\begin{equation}
\label{eq:ffield}
\gamma^\mu \nabla_\mu \psi = \lambda \psi.
\end{equation}
The mass of $\psi$ is therefore just $m = \lambda$.  The conformal
weight $h$ of a spinor field $\psi$ is given by
$h = |m| + \half$.\cite{huge} \ 
Finally recall that the gravitino $\Psi_M$ is complex, so the spinor
$\psi$ is allowed to be complex as well, and therefore represents
2 fermionic degrees of freedom on-shell.
We therefore find from~(\ref{eq:smatrix})
the following towers of spinors:
\begin{equation}
\label{eq:groupspinors}
2 \left[ (l+1,l)_{l \ge {3 \over 2}} \oplus (l+1,l)_{l \ge \ha}\right]
= 2 ({\textstyle{3 \over 2}},\half) \oplus 4(k + \half, k -\half)_{k \ge 2}.
\end{equation}

Finally we need to discuss the presence of fermionic boundary degrees of
freedom.
Although we have not carefully analysed 
the residual gauge transformations for the fermions, we can identify
the affected states by looking for vanishing equations of motion.
Vanishing equations of motion signal the presence of states which
are subject to a residual gauge symmetry \cite{pvn}.  We saw this
for the bosons at the $l{=}1$ level with
regard to the residual bosonic gauge transformations of
section~\ref{sec:resgauge}.  The vanishing of $A^m_{(lm)}$
in~(\ref{eq:smatrix}) implies through~(\ref{eq:feom}) that the equation
of motion for the associated spacetime spinor $\psi$ vanishes.  We
identify this $\psi$ as a boundary degree of freedom with
$SL(2,{\Bbb R}) \times SU(2)$ content
\begin{equation}
\label{eq:fboundary}
2({\textstyle{3 \over 2}},\half),
\end{equation}
(again, $\psi$ counts twice because it is complex).
In section~(\ref{sec:groups}) we will see that~(\ref{eq:fboundary})
falls into an $SU(1,1|2)$ multiplet with the $l{=}1$ scalars
whose equation of motion vanishes.

\subsection{Gravitini}

In two dimensions, gravitini (massless or massive)
have no on-shell
degrees
of freedom.  The story is similar to that of the graviton---in the
massless
case, the gauge symmetry ({\em i.e.} supersymmetry) eliminates the
states,
while in the massive case the equation of motion implies additional
constraints which eliminate the states.

The general theory of Kaluza-Klein compactification \cite{kksugra}
shows that massless gravitini are in one-to-one correspondence
with Killing spinors on the internal manifold, which in this case
form a ${\mathbf 2}$ of $SU(2)$.

Taking $\chi = \eta_{(\ha,m)}$ in~(\ref{eq:feom}) gives
\begin{equation}
\label{eq:iii}
\gamma^\nu \nabla_\nu \psi^\mu  - \gamma^{\mu\nu} 
\psi_\nu - \lambda \psi^\mu = 0
\end{equation}
with $\lambda = 1$.
The relation between the mass and conformal weight for a
spin-$\frac{3}{2}$
field is \cite{huge}
\begin{equation}
\label{eq:lll}
h = |m| + \ha.
\end{equation}
However, the definition of the `mass' $m$ for a spin-$\frac{3}{2}$
to use in this equation has some convention-dependence.  A careful
analysis \cite{koshelev}
\footnote{Massless Rarita-Schwinger fields were previously considered, in
the context of the AdS/CFT correspondence, in Ref.~\cite{notnastya,nastya}.}
shows that the field equation~(\ref{eq:iii})
implies that the $m$ in~(\ref{eq:lll}) should be $m = \lambda - 1$.
Indeed, with this definition of mass, gauge symmetry ({\em i.e.},
supersymmetry) is restored at $m = 0$.  To see this, consider
a supersymmetry variation $\delta \psi_\mu = \nabla_\mu
\tilde{\eta}$ where $\tilde{\eta}$ is a Killing spinor on \ad.
Using $\nabla_\mu \tilde{\eta} = \half \gamma_\mu \tilde{\eta}$,
one easily checks that~(\ref{eq:iii}) is invariant only when
$m = \lambda - 1 = 0$.

The massless gravitino, which has only boundary degrees of
freedom,
therefore has $SL(2,{\Bbb R}) \times SU(2)$ content
\begin{equation}
\label{eq:groupgravitini}
(\half,\half).
\end{equation}

\section{$SU(1,1|2)$ composition of the spectrum}
\label{sec:groups}
\setcounter{equation}{0}

In this section we show how the Kaluza-Klein modes fall into
representations of $SU(1,1|2)$, whose bosonic part is
the product of the $SU(1,1) \cong SL(2,{\Bbb R})$ isometry of \ad\ 
and the $SU(2)$ isometry of $S^2$.

The unitary irreducible representations of $SU(1,1|2)$ are well
known \cite{gst} (see \cite{deboer,huge} for 
a
modern review and applications).
The short multiplets are labelled by a half-integer ${\mathbf k}$
($k > 0$)
and have the $SL(2,{\Bbb R}) \times SU(2)$ content
\cite{huge}
\begin{equation}
\label{eq:rep}
{\mathbf k} \equiv (k,k) \oplus 2(k+\half,k-\half) \oplus (k+1,k-1).
\end{equation}
The multiplet includes four states (which are $SL(2,{\Bbb R})$
primaries), except in the case $k = \half$ where
the last state is missing,
\begin{equation}
{\mathbf \half} =  (\half,\half) \oplus 2(1,0).
\end{equation}
There are no singleton representations \cite{deboer}.

We now assemble the $SU(1,1|2)$ multiplets
from~(\ref{eq:groupbosons}),~(\ref{eq:groupgauge}),~(\ref{eq:fboundary}) and
(\ref{eq:groupspinors}), and~(\ref{eq:groupgravitini}).
From the massless graviton, $l=0$ vector and gravitino we form
a single ${\mathbf \half}$ of $SU(1,1|2)$.
These fields are just the restriction to the lowest $S^2$ harmonic
of the original fields $(g_{MN}, \Psi_M, A_M)$.  They are pure gauge
modes in two dimensions.
From the infinite towers of scalars, vectors, and spinors we have
two towers
$2 ({\mathbf k})_{k \ge 2}$ of modes with physical degrees of
freedom in the bulk of \ad.
Finally we see that
the scalar mode with undetermined $h$ in~(\ref{eq:groupgauge})
must be $(1,1)$ in order to fit the remaining modes into
$SU(1,1|2)$ representations.  Assembling the remaining modes gives
$2 ({\mathbf 1})$.
One of these ${\mathbf 1}$'s contains fields whose equations of motion
were found to vanish, signaling the presence of residual gauge
symmetries which eliminate the bulk degrees of freedom of these modes.
The other ${\mathbf 1}$ contains a bulk scalar, 
a complex bulk spinor, and a triplet of massless vector fields
transforming in the ${\mathbf 3}$ of $SU(2)$.
Because we only linearized the equations, we could not see the nonabelian
gauge symmetry presumably possessed by these vectors.

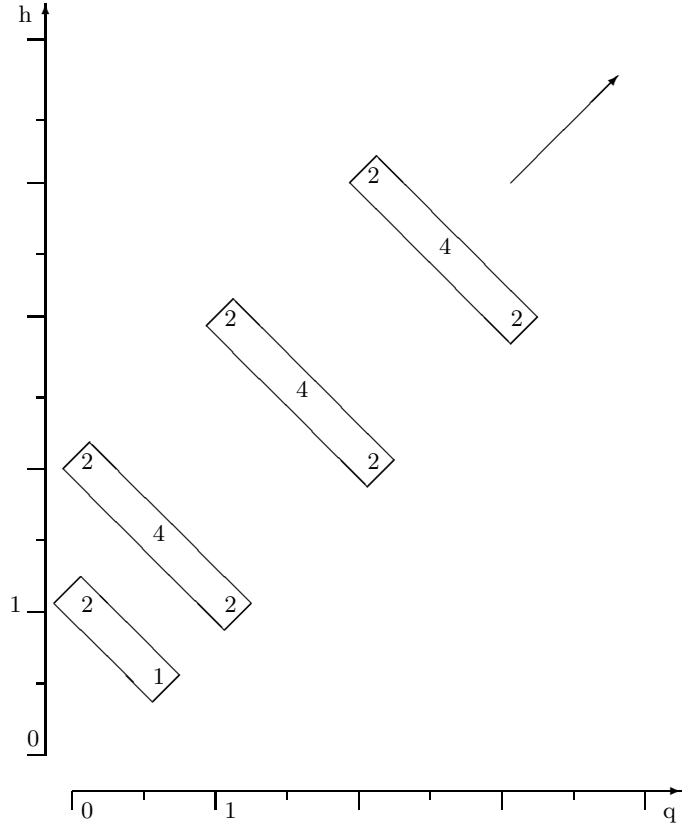
\begin{figure}[tbh]
\begin{center}
\setlength{\unitlength}{2960sp}%
\begingroup\makeatletter\ifx\SetFigFont\undefined%
\gdef\SetFigFont#1#2#3#4#5{%
  \reset@font\fontsize{#1}{#2pt}%
  \fontfamily{#3}\fontseries{#4}\fontshape{#5}%
  \selectfont}%
\fi\endgroup%
\begin{picture}(5637,6877)(601,-9326)
\thinlines
\put(2251,-5161){\line( 1,-1){1350}}
\put(3601,-6511){\line( 1, 1){225}}
\put(3826,-6286){\line(-1, 1){1350}}
\put(2476,-4936){\line(-1,-1){225}}
\put(2401,-5161){\makebox(0,0)[lb]{\smash{\SetFigFont{9}{10.8}{\rmdefault}{\mddefault}{\updefault}2}}}
\put(3001,-5761){\makebox(0,0)[lb]{\smash{\SetFigFont{9}{10.8}{\rmdefault}{\mddefault}{\updefault}4}}}
\put(3601,-6361){\makebox(0,0)[lb]{\smash{\SetFigFont{9}{10.8}{\rmdefault}{\mddefault}{\updefault}2}}}
\put(3451,-3961){\line( 1,-1){1350}}
\put(4801,-5311){\line( 1, 1){225}}
\put(5026,-5086){\line(-1, 1){1350}}
\put(3676,-3736){\line(-1,-1){225}}
\put(3601,-3961){\makebox(0,0)[lb]{\smash{\SetFigFont{9}{10.8}{\rmdefault}{\mddefault}{\updefault}2}}}
\put(4201,-4561){\makebox(0,0)[lb]{\smash{\SetFigFont{9}{10.8}{\rmdefault}{\mddefault}{\updefault}4}}}
\put(4801,-5161){\makebox(0,0)[lb]{\smash{\SetFigFont{9}{10.8}{\rmdefault}{\mddefault}{\updefault}2}}}
\put(1051,-6361){\line( 1,-1){1350}}
\put(2401,-7711){\line( 1, 1){225}}
\put(2626,-7486){\line(-1, 1){1350}}
\put(1276,-6136){\line(-1,-1){225}}
\put(1201,-6361){\makebox(0,0)[lb]{\smash{\SetFigFont{9}{10.8}{\rmdefault}{\mddefault}{\updefault}2}}}
\put(1801,-6961){\makebox(0,0)[lb]{\smash{\SetFigFont{9}{10.8}{\rmdefault}{\mddefault}{\updefault}4}}}
\put(2401,-7561){\makebox(0,0)[lb]{\smash{\SetFigFont{9}{10.8}{\rmdefault}{\mddefault}{\updefault}2}}}
\put(901,-8761){\line(-1, 0){150}}
\put(901,-8161){\line(-1, 0){ 75}}
\put(901,-6961){\line(-1, 0){ 75}}
\put(901,-5761){\line(-1, 0){ 75}}
\put(901,-4561){\line(-1, 0){ 75}}
\put(901,-3436){\line(-1, 0){ 75}}
\put(901,-7561){\line(-1, 0){150}}
\put(901,-6361){\line(-1, 0){150}}
\put(901,-5086){\line(-1, 0){150}}
\put(901,-3961){\line(-1, 0){150}}
\put(901,-2761){\line(-1, 0){150}}
\put(901,-8761){\vector( 0, 1){6300}}
\put(4801,-3961){\vector( 1, 1){900}}
\put(1126,-9061){\vector( 1, 0){5100}}
\put(1126,-9061){\line( 0,-1){150}}
\put(1726,-9136){\line( 0, 1){ 75}}
\put(2326,-9061){\line( 0,-1){150}}
\put(2926,-9061){\line( 0,-1){ 75}}
\put(3526,-9061){\line( 0,-1){150}}
\put(4126,-9061){\line( 0,-1){ 75}}
\put(4726,-9061){\line( 0,-1){150}}
\put(5326,-9061){\line( 0,-1){ 75}}
\put(5926,-9061){\line( 0,-1){150}}
\put(976,-7486){\line( 1,-1){825}}
\put(1801,-8311){\line( 1, 1){225}}
\put(2026,-8086){\line(-1, 1){825}}
\put(1201,-7261){\line(-1,-1){225}}
\put(601,-7561){\makebox(0,0)[lb]{\smash{\SetFigFont{9}{10.8}{\rmdefault}{\mddefault}{\updefault}1}}}
\put(751,-8686){\makebox(0,0)[lb]{\smash{\SetFigFont{9}{10.8}{\rmdefault}{\mddefault}{\updefault}0}}}
\put(676,-2611){\makebox(0,0)[lb]{\smash{\SetFigFont{9}{10.8}{\rmdefault}{\mddefault}{\updefault}h}}}
\put(1201,-9286){\makebox(0,0)[lb]{\smash{\SetFigFont{9}{10.8}{\rmdefault}{\mddefault}{\updefault}0}}}
\put(2401,-9286){\makebox(0,0)[lb]{\smash{\SetFigFont{9}{10.8}{\rmdefault}{\mddefault}{\updefault}1}}}
\put(6076,-9286){\makebox(0,0)[lb]{\smash{\SetFigFont{9}{10.8}{\rmdefault}{\mddefault}{\updefault}q}}}
\put(1201,-7561){\makebox(0,0)[lb]{\smash{\SetFigFont{9}{10.8}{\rmdefault}{\mddefault}{\updefault}2}}}
\put(1801,-8161){\makebox(0,0)[lb]{\smash{\SetFigFont{9}{10.8}{\rmdefault}{\mddefault}{\updefault}1}}}
\end{picture}
\end{center}
\caption{A state at a given value of $q$ forms a ${\mathbf 2q + 1}$ of $SU(2)$.
The top element of the ${\mathbf 2 q + 1}$ has $J_0^3 = +q$ and the bottom
element of $J_0^3 = - q$.  Those states with $h = J_0^3$ are chiral
primaries and those with $h = - J_0^3$ are chiral anti-primaries.}
\end{figure}

\section{Conclusion and Discussion}
\setcounter{equation}{0}

We have determined the Kaluza-Klein spectrum of ${\mathcal{N}}{=}2$, $D{=}4$
supergravity on \adst.
The $SU(1,1|2)$ composition of the spectrum is
${\mathbf \half} \oplus 2({\mathbf k})_{{\Bbb Z} \ni k \ge 1}$.
Furthermore we have categorized the modes according to whether they
have on-shell degrees of freedom in the bulk of \ad\ or whether,
because of residual gauge transformations, they may be gauged away
everywhere in the bulk of \ad\ and hence only represent boundary
degrees of freedom.

In two dimensions one also encounters modes without any gauge
symmetry yet which have no on-shell degrees of freedom in the
bulk, such as gravitons and gravitini.
We showed explicitly that the equations of motion for the massive
gravitons implied that they have no independent degrees of freedom.
Our analysis for the gravitini was less careful, but the result must
follow by supersymmetry.  It is conceivable that one could resurrect
these modes
(or any other modes one fancies)
on the boundary
by adding suitable boundary terms to the supergravity
action.  However they are not required, and indeed our $SU(1,1|2)$
analysis has not found any modes lacking.

\hbox{}

\noindent As this work was near completion we learned of similar investigations
by J.\ Lee and S.\ Lee~\cite{lee}. 

\section*{Acknowledgements}
\setcounter{equation}{0}

It is a pleasure to thank R.\ Britto-Pacumio, F.\ Larsen, A.\ Lawrence, S.\
Lee, J.\
Maldacena, A.\ Strominger and
A.\ Volovich for
helpful discussions.
This work was supported by an NSERC PGS B Scholarship, an NSF graduate
fellowship and 
DOE grant DE-FG02-91ER40654.

\appendix

\section{Conventions}
\setcounter{equation}{0}

We use the mostly positive metric signature, $\eta_{MN}
= {\rm diag}(-1,+1,+1,+1)$.
Upper case Latin indices
$M,N,\ldots = 0,1,2,3$ run over \adst,
lower case Latin indices
$m,n = 2,3$
are internal indices and run over $S^2$,
and Greek indices $\mu,\nu = 0,1$ denote \ad\ spacetime indices.
We use $x^\mu$ for coordinates on $AdS_2$ and $y^m$ for coordinates
on $S^2$.
The Levi-Civita symbol $\epsilon$ is a
true tensor, with $\epsilon_{0123} = \sqrt{-g}$, and $*F^{MN} = \ha
\epsilon^{MNPQ} F_{PQ}$.
Curly braces around a pair of indices indicate symmetrization with
the
trace removed, {\it i.e.}
\begin{equation}
h_{\{ mn \}} = \ha ( h_{mn} + h_{nm} - g_{mn} {h^p}_p).
\end{equation}

Let $\gamma^\mu$ and $\gamma^m$ respectively satisfy the algebras
$\{ \gamma^\mu, \gamma^\nu \} = 2 g^{\mu \nu}$ and $\{ \gamma^m,
\gamma^n \} = 2 g^{mn}$.
Multiple indices denote antisymmetrization with unit
weight, {\it e.g.}
$\gamma^{mn} = \ha(\gamma^m \gamma^n-\gamma^n\gamma^m)$.
The chirality operator on $S^2$ is given by $\gamma = -\frac{i}{2}
\epsilon_{mn} \gamma^{mn}$.
A representation of the four-dimensional Dirac algebra $\{
\Gamma^M, \Gamma^N \} = 2 g^{MN}$ is then furnished by
$\Gamma^\mu = \gamma^\mu \otimes \gamma$ and $\Gamma^m = {\mathbf 1}_2
\otimes \gamma^m$.  In the text we will frequently omit the tensor
product symbol and write simply $\Gamma^\mu = \gamma^\mu
\gamma$ and $\Gamma^m = \gamma^m$.
The four-dimensional chirality operator is $\Gamma^5 = -\frac{i}{4!}
\epsilon_{MNPQ} \Gamma^{MNPQ} = \ha
\epsilon_{\mu \nu} \gamma^{\mu \nu}
\gamma$ (omitting the tensor product symbol).


\newpage

\begin{thebibliography}{99}

\bibitem{juan} J.\ Maldacena, {\em The Large $N$ Limit of
Superconformal Field Theories and Supergravity}, Adv.\ Theor.\ Math.\
Phys.\ {\bf 2} (1998) 231--252, {\tt hep-th/9711200}.

\bibitem{gkp} S.\ S.\ Gubser, I.\ R.\ Klebanov and A.\ M.\ Polyakov,
{\em Gauge Theory Correlators from Non-Critical String Theory},
Phys.\ Lett.\ {\bf B428} (1998) 105--114,
{\tt hep-th/9802109}.

\bibitem{witten} E.\ Witten, 
{\em Anti de Sitter Space and Holography}, Adv.\ Theor.\ Math.\ Phys.\ {\bf
2} (1998) 253--291,
{\tt hep-th/9802150}.

\bibitem{huge} O.\ Aharony, S.\ S.\ Gubser, J.\ Maldacena, H.\ Ooguri and
Y.\ Oz, {\em Large $N$ Field Theories, String Theory and Gravity},
CERN-TH/99-122, HUTP-99/A027, LBNL-43113, RU-99-18, UCB-PTH-99/16,
{\tt hep-th/9905111}.

\bibitem{sezgin} S.\ Deger, A.\ Kaya, E.\ Sezgin and P.\ Sundell, {\em
Spectrum of $D=6$, $N=4$b Supergravity on $AdS_3\times S^3$},
Nucl.\ Phys.\ {\bf B536} (1998) 110--140,
{\tt hep-th/9804166}.

\bibitem{deboer} J.\ de Boer, {\em Six-Dimensional Supergravity
on $S^3 \times AdS_3$ and 2d Conformal Field Theory},
Nucl.\ Phys.\ {\bf B548} (1999) 139--166,
{\tt hep-th/9806104}.

\bibitem{larsen} F.\ Larsen, {\em The Perturbation Spectrum of Black
Holes in $N = 8$ Supergravity}, Nucl.\ Phys.\ {\bf B536} (1998) 258--278,
{\tt hep-th/9805208}.

\bibitem{fujii2} A.\ Fujii, R.\ Kemmoku and S.\ Mizoguchi, {\em $D = 5$
Simple Supergravity on $AdS_3 \times S^2$ and $N = 4$ Superconformal
Field Theory}, {\tt hep-th/9811147}.

\bibitem{fujii3} A.\ Fujii and R.\ Kemmoku, {\em $D = 5$ Simple
Supergravity on $AdS_2 \times S^3$},
Phys.\ Lett.\ {\bf B459} (1999) 137--144,
{\tt hep-th/9903231}.

\bibitem{frag} J.\ Maldacena, J.\ Michelson and A.\ Strominger, {\em
Anti-de Sitter Fragmentation}, JHEP {\bf 02} (1999) 011, {\tt hep-th/9812073}.

\bibitem{andy} A.\ Strominger, {\em $AdS_2$ Quantum Gravity and String
Theory}, JHEP {\bf 01} (1999) 007, {\tt hep-th/9809027}.

\bibitem{gt}
G.\ W.\ Gibbons and P.\ K.\ Townsend, {\em Black holes and Calogero
models},
Phys.\ Lett.\ {\bf B454} (1999) 187--192,
{\tt hep-th/9812034}.

\bibitem{krv} H.\ J.\ Kim, L.\ J.\ Romans and P.\ van Nieuwenhuizen, {\em
Mass Spectrum of Chiral Ten-Dimensional $N=2$ Supergravity on $S^5$},
Phys.\ Rev.\ {\bf D32} (1985) 389--399.

\bibitem{duff} M.\ J.\ Duff, {\em Anti-de Sitter Space, Branes,
Singletons, Superconformal Field Theories and All That},
Int.\ J.\ Mod.\ Phys.\ {\bf A14} (1999) 815--844,
{\tt hep-th/9808100}.

\bibitem{freedas} D.\ Z.\ Freedman and A.\ Das, {\em Gauge Internal
Symmetry in Extended Supergravity}, Nucl.\ Phys.\ {\bf B120} (1977)
221--230.

\bibitem{freund} P.\ G.\ O.\ Freund and M.\ A.\ Rubin, {\em Dynamics of
Dimensional Reduction}, Phys.\ Lett.\ {\bf B97} (1980) 233--235.

\bibitem{fujii} Y.\ Fujii and K.\ Yamagishi, {\em Killing Spinors on
Spheres and Hyperbolic Manifolds}, J.\ Math.\ Phys.\ {\bf 27} (1986)
979--981.

\bibitem{lpr} H.\ L\"u, C.\ N.\ Pope and J.\ Rahmfeld, {\em A Construction
of Killing Spinors on $S^n$}, 
CTP~TAMU-22/98, LPTENS-98/22, SU-ITP-98/31,
{\tt hep-th/9805151}.

\bibitem{mathews} J.\ Mathews, {\em Tensor Spherical Harmonics},
California Institute of Technology, 1981.

\bibitem{polishchuk} A.\ Polishchuk, {\em Massive Symmetric Tensor
Field on $AdS$},
JHEP {\bf 07} (1999) 007,
{\tt hep-th/9905048}.

\bibitem{cenr} A.\ Casher, F.\ Englert, H.\ Nicolai and M.\ Rooman, {\em
The Mass Spectrum of Supergravity on the Round Seven-Sphere}, Nucl.\ 
Phys.\ {\bf B243} (1984) 173--188.

\bibitem{pvn} P.\ van Nieuwenhuizen, {\em The Complete Mass Spectrum
of $d = 11$ Supergravity Compactified on $S_4$ and a General Mass
Formula for Arbitrary Cosets $M_4$}, Class.\ Quant.\ Grav.\ {\bf
2} (1985) 1--20.

\bibitem{kksugra} M.\ J.\ Duff, B.\ E.\ W.\ Nilsson and C.N.\ Pope, {\em
Kaluza-Klein Supergravity}, Phys.\ Rep.\ {\bf 130} (1986)~1--142.

\bibitem{koshelev} A.\ S.\ Koshelev and O.\ A.\ Rythkov, {\em Note on the
Massive Rarita-Schwinger FIeld in the AdS/CFT Correspondence},
Phys.\ Lett.\ {\bf B450} (1999) 368--376,
{\tt hep-th/9812238}.

\bibitem{notnastya} S.\ Corley, {\em The Massless Gravitino and the AdS/CFT
Correspondence}, Phys.\ Rev.\ {\bf D59} (1999) 086003, {\tt hep-th/9808184}.

\bibitem{nastya} A.\ Volovich, {\em Rarita-Schwinger Field in the AdS/CFT
Correspondence}, JHEP {\bf 09} (1998) 022, {\tt hep-th/9809009}.

\bibitem{gst} M.\ G\"unaydin, G.\ Sierra and P.\ K.\ Townsend, {\em
The Unitary Supermultiplets of $d = 3$ Anti-de Sitter and $d = 2$
Conformal Superalgebras}, Nucl.\ Phys.\ {\bf B274} (1986) 429--447.

\bibitem{sugra} P.\ van Nieuwenhuizen, {\em Supergravity}, Phys.\ Rep.\
{\bf 68} (1981) 189--398.

\bibitem{lee}
J.\ Lee and S.\ Lee, {\em Mass Spectrum of $D=11$ Supergravity
on $AdS_2 \times S_2 \times T^7$},
KIAS-P99038,
{\tt hep-th/9906105};
S.\ Lee at the
Carg\`{e}se~'99 ASI on Progress in String Theory and M-Theory
``Gong Show''.

\end{thebibliography}
\end{document}